# ASAP (Automatic Software for ASL Processing): A toolbox for processing Arterial Spin Labeling images

Virginia Mato Abad[1], Pablo García-Polo[2], Owen O'Daly[3], Juan Antonio Hernández-Tamames[1], Fernando Zelaya[3]

[1]Laboratorio de Análisis de Imagen Médica y Biometría (LAIMBIO), Universidad Rey Juan Carlos, Mostoles, Madrid, Spain

[2]M+Visión Advanced Fellowship, Medical Imaging Lab., Hospital Universitario de Fuenlabrada, Fuenlabrada, Madrid, Spain

[3]Department of Neuroimaging, Institute of Psychiatry, King's College London, London, United Kingdom

**Corresponding Author:**

Virginia Mato Abad
Universidad Rey Juan Carlos
Departamental II. Despacho 157.
Campus de Móstoles, C/Tulipán s/n
28933, Móstoles, Madrid (Spain)
Telephone: +34 914888522
virginia.mato@urjc.es

**Abstract**

The method of Arterial Spin Labeling (ASL) has experienced a significant rise in its application to functional imaging, since it is the only technique capable of measuring blood perfusion in a truly non-invasive manner. Currently, there are no commercial packages for processing ASL data and there is no recognised standard for normalising ASL data to a common frame of reference. This work describes a new Automated Software for ASL Processing (ASAP) that can automatically process several ASL datasets. ASAP includes functions for all stages of image pre-processing: quantification, skull-stripping, co-registration, partial volume correction and normalization. To assess the applicability and validity of the toolbox, this work shows its application in the study of hypoperfusion in a sample of healthy subjects at risk of progressing to Alzheimer's Disease. ASAP requires limited user intervention, minimising the possibility of random and systematic errors, and produces cerebral blood flow maps that are ready for statistical group analysis. The software is easy to operate and results in excellent quality of spatial normalisation. The results found in this evaluation study are consistent with previous studies that find decreased perfusion in Alzheimer's patients in similar regions and demonstrate the applicability of ASAP.

**Keywords**: Arterial Spin Labeling, Cerebral Blood Flow, Automatic Processing, Partial volume effect, Alzheimer's Disease

**1. Introduction**

Arterial Spin Labelling (ASL) has become a popular magnetic resonance technique for imaging brain function. It is entirely non-invasive and capable of quantitatively determining regional blood perfusion; providing therefore a significant advantage over contrast agent based methods like $^{15}O$ enriched $H_2O$ Positron Emission Tomography (PET) or Gadolinium-based Dynamic Susceptibility Contrast Magnetic Resonance Imaging (DSC-MRI). The basic principle of ASL is to employ arterial blood water itself as contrast agent to measure perfusion. For cerebral blood flow (CBF) this is obtained by tagging a bolus of arterial blood in the region of the carotid arteries. The magnetization of inflowing blood water protons is inverted in that region by means of an external radiofrequency pulse, which is applied either as a short pulse (10-20ms) or as a continuous or pseudo-continuous burst of radiofrequency (1-2s) in the presence of a gradient. After a period of time (post-labelling delay), blood labelled with inverted signal is delivered to the entire brain through the smaller arteries and capillaries. This labelled arterial blood signal gives rise to a reduction in the image intensity when compared to a non-labelled (control) image. The control and labelled images are subtracted to generate a 'perfusion weighted' image. The intensity of each voxel will reflect the amount of arterial blood delivered in the inversion time; and through the use of a suitable model, the difference image is transformed to a map of CBF in conventional physiological units of ml blood/100g tissue/min.

The availability of ASL as a routine method for assessment of basal CBF data has provided the possibility to examine brain physiology and generate a marker to probe functional differences between groups. ASL is increasingly used in clinical studies of cerebral perfusion and has shown its validity in measuring perfusion changes in several neurodegenerative diseases including Alzheimer Disease (AD) [1,2]; as well as in psychiatric studies [3], pharmacology [4] and pain [5]. However, to perform this type of analysis, multiple image processing steps are required: quantification, registration, normalization to a standard space, partial volume correction, etc.

Partial volume effects (PVE) are a consequence of limited spatial resolution in imaging and especially in ASL, where the low signal-to-noise (SNR) ratio leads to the need to employ larger voxels. In an effort to increase SNR, tissue specific saturation pulses are applied to the volume of interest to suppress the static tissue signal  This is known as 'background suppresion' and it is now used extensively in ASL [6]. Nevertheless, the change in the received signal due to blood water proton relaxation remains very small, such that voxels are typically of the order of 3x3x6mm, generating the need to employ some form of PVE correction as each voxel is likely to contain signal mixing from different tissue types. Normal grey matter (GM) perfusion values are around 60ml/100g/min while white matter (WM) values are significantly lower (20ml/100g/min) [7]. Due to the relative insensitivity of ASL in white matter, the prime interest when using this technique is the study of pure GM perfusion.  However, in voxels containing (for example) 50% GM and 50% WM, the CBF values could be underestimated by up to one-third. PVE is of paramount importance in the study of neurodegenerative diseases where GM atrophy significantly affects CBF quantification and therefore the comparison of patient data with control populations.

The absence of a standard approach for data processing has been partly driven by the fact that several ASL methodologies have evolved independently [8]. Therefore, there is no recognised standard for normalising ASL data to a common frame of reference. This lack of a harmonised processing pipeline contributes to the potential discrepancies in studies of brain perfusion across different laboratories [9].

A number of packages, such as BASIL [10] and ASLTbx [11] provide a set of functions for pre-processing of ASL data and they both are free for academic use. BASIL consists of a collection of tools from the Functional Software Library (FSL) suite [12] that aid in the quantification and subsequent spatial processing of CBF images acquired with ASL. BASIL is based on Bayesian inference principles and was originally developed for ASL data acquired with several post-labelling delays (known as 'multi-TI' data). ASLTbx is a MATLAB [13] and SPM [14] based toolkit for processing ASL data, which requires basic MATLAB script programming.

These packages typically perform a step-by-step and subject-by-subject processing and require a large amount of manual operation. To date, a toolbox supporting a fully automated processing of raw ASL data, with minimum user intervention that can be used for effective comparison of group data, is not yet available.

In this article, we describe the development, implementation and test of an ASL processing toolbox (ASAP) that can automatically process several ASL datasets, from their raw image format to a spatially normalised, smoothed (if desired) version, with minimal user intervention. Ease of operation has been facilitated by a graphical user interface (GUI) whose operation is entirely intuitive.  After the user sets the input/output and processing parameters using the GUI, the toolbox fully executes all processing steps for datasets of any number of subjects and results in data ready for second level statistical analysis. The data can be written in a variety of formats to facilitate its inclusion in several software packages for group analysis. The toolbox also has a facility to display the spatially normalised data in a manner that facilitates quality control by the user.

To assess the applicability and validity of the toolbox, we demonstrate its use in the study of hypoperfusion in a sample of healthy subjects at risk of progressing to Alzheimer's Disease (AD).

## 2. Methods

### 2.1. Toolbox processing procedures

ASAP has been developed in MATLAB with the goal of simplifying the process of quantification and pre-processing of ASL studies. It includes functions like CBF quantification, skull stripping, co-registration, partial volume correction and normalisation. Different processing strategies have been made available depending on user requirements:

- System requirements: ASAP is written in MATLAB under a Unix system (Linux or Mac OS) but it is not entirely a stand-alone utility. It accesses both FSL software and SPM libraries, which are two of the most widely available image processing platforms for MRI. These are invoked by the toolbox and are transparent to the user, but they must be installed independently by each user and added to the MATLAB path (including the FSLDIR environment variable). The software works equally well with earlier version of SPM or with the latest release (SPM-12).

- Input data: The ASL input data can be the raw difference image (control image – labelled image) or the perfusion image (CBF map). Regardless of the input or the ASL modality used, computation of the CBF map is made according to the formula proposed in the recent article "Recommended implementation of arterial spin-labeled perfusion MRI for clinical applications" published by Alsop et al [15].  For subsequent spatial co-registration and normalisation, the user is able to choose between providing a high-resolution T1-weighted or T2-weighted structural scan. DICOM, NIfTI or ANALYZE formats are accepted.

- Resolution: The user can select between two different execution methods regarding the resolution of the images: the low-resolution native space of ASL or up-sampling the ASL images to the structural image high-resolution grid, typically of the order of 1x1x1 mm voxel size (acquisition matrix of 288x288 or 512x512 voxels with full brain coverage. The up-sampling is made by means of the spatial interpolation 'Nearest Neighbour', which preserves the grey values of the original voxel and ensures the consistency of CBF values. After the spatial normalization, the ASL voxel size is 2x2x2 mm, the resolution of the MNI template.

- Cerebral blood flow quantification: Due to the fact that most multi-TI ASL sequences are currently only available as experimental or prototype versions, the toolbox only includes CBF quantification for single inversion time data. In that case, the ASL difference image should be provided as input. The CBF quantification map is calculated using the formula currently recommended method [15]. In addition to the difference image, a reference proton density image and the post labelling delay time employed are also required.

- Partial volume correction (PVC): ASAP provides the option of PVC of the ASL data. In its current version, two different methods are provided: 1) the method described by Asllani [16] and 2) a method based on a previous approached developed for PET (from here referred to as 'tghe PET method') that assumes perfusion of WM is globally 40% of that of GM for correction of resting CBF [17]. Although the later is a more simplistic approach and has been largely superseded by the methods introduced by Asllani and Chappell, this method (hereafter referred to as the PET correction) is available in our toolbox because it has been applied historically in earlier  ASL studies [18-20]. Asllani's algorithm is based on linear regression and represents the voxel intensity as a weighted sum of pure tissue contribution, where the weighting coefficients are the tissue's fractional volume in the voxel. This algorithm is able to estimate the CBF for grey matter (GM) and white matter (WM) independently. The PET correction assumed that all contributions to perfusion are from brain tissue and that cerebrospinal fluid has no contribution. In that case, ASL intensities are corrected according to the following equation:

$$I_{corr}=I_{uncorr}/(P_{GM}+0.4*P_{WM})$$

where $I_{corr}$ and $I_{uncorr}$ are the corrected and uncorrected intensities, the 0.4 factor is the global ratio between WM and GM and $P_{GM}$ and $P_{WM}$ are the probabilities of GM and WM, respectively. The PVC option is only available when working in the low-resolution ASL space, thus having co-registered the high-resolution structural image to the ASL image.

● Execution mode: The toolbox includes a Graphical User Interface (GUI) where all the input data can be setup manually. Also, it has a batch mode for advanced users.

The main procedure of ASAP is shown in Figure 1 and includes the following steps:

1. Optional CBF quantification for pCASL and PASL sequences.
2. Reorient the images. Structural and ASL images are reoriented to the AC-PC plane (Anterior Commissure - Posterior Commissure) and their origins are set to the AC. Setting of a common origin is advisable for superior performance of the subsequent processing steps. If the PD image is available, the PD image is reoriented to the AC-PC plane, applying the same transformation to the ASL image.
3. Rough skull-stripping of the initial resting state ASL map using the FSL Brain Extraction Tool (bet) using a conservative threshold. This step is useful for noisy ASL maps, in order to increase the quality of the rigid co-registration with the structural scan.
4. Estimation of the brain mask. Brain mask from the structural volume can be calculated by two different options: the FSL bet tool (recommended for T2-weighted high-resolution scan) or the SPM segmentation task (recommended for T1-weighted high resolution scan). The brain mask is required for excluding out-of-brain voxels, often encountered in subtraction techniques such as ASL. Segmentation of GM and WM probability maps is also required for the partial volume correction step.
5. Rigid co-registration between ASL and structural images using SPM function. ASL images are normally co-registered to anatomical images so they can be later normalized to the MNI space (or any other standard space) for group analysis. Also, the co-registration is required for the partial volume correction. T1-weighted or T2-weighted images can be used for co-registration. If direct co-registration of ASL and structural images is not reliable because of the poor signal-to-noise ratio and the limited structural features of perfusion images, the proton density (PD) image can also be used for co-registration, moving the ASL data in the process. Depending on the selected resolution, the co-registration will be made in the native space of the ASL data (down-sampling the resolution of the structural scan) or up-sampling the ASL to the high-resolution of the structural volume by interpolation.
6. Partial Volume Correction of the ASL maps using the methods available. Information

about the proportion of each tissue type (grey matter, white matter, and cerebrospinal fluid) is used to correct perfusion data. The method described by Asllani estimates both, partial GM and partial WM ASL maps. The PET correction method only estimates the partial GM ASL map. This option is only available if the structural scan has been down-sampled by means of the rigid co-registration step to the ASL image.

7.  Skull-stripping of the ASL data. Apply the brain mask previously calculated to the co-registered and partial volume corrected ASL maps in order to exclude artefactual, finite 'perfusion' values in the extra-cerebral space (These arise in all ASL modalities because of the subtraction of control and labelled images).

8.  Spatial normalization. For comparison across subjects, location correspondence has to be established, so registration of all the individual images to a standardized space is required. Here, the images (both ASL and structural) are normalized to the MNI standard space using: 1) a MNI template selected by the user or 2) the transformation matrix earlier calculated by SPM during the segmentation process.

9.  Smoothing. The resultant images in the standard space are ready for voxel-based statistical analysis. However, these images are commonly multiplied by a smoothing kernel larger than the voxel dimension to satisfy the random-field approximation employed in parametric statistics. The SPM Gaussian smoothing kernel is applied to the final ASL maps, the size of the kernel (in mm) is selectable by the user.

10. The resultant images can be directly used for statistical analysis. This procedure is very flexible, as most of the steps are optional. Thus, users can freely design the pipeline that best fit their needs.

## 2.2. Testing the hypoperfusion in healthy subjects in risk of developing Alzheimer's disease by using the toolbox

Several studies [1,21-25] have shown that Alzheimer's patients suffer from decreased perfusion in specific cortical and sub-cortical areas that may be associated to the subsequent cognitive and structural degeneration. A subgroup of the "Proyecto Vallecas" study, a 4-year longitudinal study over 1,000 subjects to assess normal healthy ageing and the appearance of neurodegenerative diseases, in particular AD; was selected to validate this hypothesis and demonstrate ASAP.

### 2.2.1. Subjects

A two-group study comparing 25 healthy elderly subjects (7 men and 18 women, mean age 75±3.6 years) and 25 elderly subjects at risk of developing Alzheimer's disease (8 men and 17 women, mean age 77±4.5 years) was performed. All subjects were first included in the study as healthy subjects based on several psychological and neurological tests, including the Geriatric Depression Scale [26], a Mini–Mental State Examination (MMSE) [27] above 24 and Functional Activities Questionnaire (FAQ) [28] scores above 6 at the baseline assessment. All subjects included in the study show no signs of dementia or severe cognitive deterioration and they are able to manage and independent life without any mental disorder (cognitive or psychiatric) impeding daily functioning. All subjects underwent MRI examination as well as psychological and neurological assessment every 6-12 months. Informed consent was obtained from all participants prior to evaluation. The subjects selected as subjects at risk of developing AD were those whose left and right hippocampi suffered from a volume loss greater than 2 standard deviations from the sample mean.

## 2.2.2. Acquisition

All subjects underwent MRI examination in a 3T Signa HDx MR scanner (GE Healthcare, Waukesha, WI) using an eight-channel phased array coil. The first sequence was a 3D T1 weighted SPGR with a TR=10.024ms, TE=4.56ms, TI=600ms, NEX=1, acquisition matrix=288x288, full brain coverage, resolution=1x1x1mm, flip angle=12. The second sequence was a 3D pCASL pulse sequence with full brain coverage, matrix size= 128x128, resolution=1.875x1.875x4mm, flip angle = 155, labelling time 1.5s, post-labelling delay=2.025s, TR=4.733s, TE=9.812ms, NEX=3, acquisition time ~6min and was used to generate the regional cerebral blood flow (rCBF) maps. Both the perfusion difference image and the proton density image produced by this sequence were available for the study.

## 2.2.3. Image processing

All 3D T1 weighted images were processed with Freesurfer [29] in order to obtain the cortical and subcortical volumes for each subject. The left and right hippocampi volume (LHV, RHV) were normalised by the total GM volume. This normalised measure allowed us to divide the sample into three groups: Control group ([LHV, RHV]>(mean hippocampus (MH)+1std.)), mean group (MH-2std.<[LHV, RHV]<MH+1std.) and probable AD group (PAD) ([LHV, RHV]<(MH-2std.)). A selection of 25 PAD subjects and 25 age and gender matched controls was the final subset used to validate ASAP.

Thus, the input images for each subject for ASAP were two DICOM series: a 3D T1 weighted image and a raw ASL sequence (control-labelled subtraction and proton density images). The resulting processing pipeline (as described above) is shown in Figure 1. To evaluate the effect of PVE correction, prior to MNI normalization two different options were applied to the perfusion maps: no PVE correction and the Asllani's PVE correction with a regression-kernel of size 5x5x1voxels. Therefore, for each subject, original CBF maps and Asllani's PVE-corrected CBF maps were obtained.

## 2.2.4. Statistical analysis

Normalized and smoothed (6mm Gaussian kernel) CBF maps produced by ASAP (both PVE corrected and uncorrected) were employed for the voxel-based statistical analysis. Statistical maps for rejecting the null hypothesis of equal perfusion between healthy and subjects at risk of developing AD were generated by means of a two sample t-test analysis within the General Linear Model (GLM) (with gender and age as covariates and mean CBF value of each subject as a regressor) as implemented in the SPM software suite.

### 3. Results

### 3.1. A MATLAB Toolbox for processing ASL images: ASAP

ASAP has been developed for fully automated processing of ASL data. It is an open-source package and is freely available (*sites.google.com/site/asltoolbox*). ASAP provides a user friendly Graphical User Interface GUI (Figure 2). Users can perform several interactions with the embedded functions, e.g., setting inputs, outputs or different processing parameters. In the "Input Files" panel (see Figure 2) users can select all the input data while the "Output Directories" panel is used to designate the directory where the output files will be saved. In the "Options" panel, users can select the different processing parameters.

In addition, the advanced mode includes a "load batch files" option for loading the input files from text files to avoid having to select the input data individually through the GUI. With this option, a large number of datasets can be loaded into the toolbox for subsequent processing using the "Options" set in the panel and the same options will apply throughout for all subjects. In addition, the advanced mode contains the 'ROI Statistics' GUI (Figure 3) that offers the option to extract CBF values from anatomically or functionally defined Regions of Interest (ROI). This facility can simultaneously extract mean, median and maximum values from several ROIs in several CBF maps. Users only have to select the input files, ASL data and ROI masks (NIfTI or .mat files are accepted), through the GUI ("Select files" action) or in batch mode ("Load files" action) from text files. Output results are saved in text files that can easily be incorporated into statistical analysis packages such as SPSS, etc.

Resultant files are stored in the directories specified by the user. Each procedure of the pipeline produces a new file, every step is recorded and files are not overwritten. The MNI normalized images can be directly used for statistical analysis, however, users can also use the intermediary results. As stated before, most of the steps described above are optional, so the procedure is very flexible and users can freely design the most appropriate pipeline. Also, the toolbox is designed to aid in reproducing some analysis by avoiding some processing steps. This feature is useful, for example, for applying different methods for partial volume correction on the same input data: if there are GM, WM and CSF maps in the same directory as the input structural scan, the toolbox does not apply the SPM segmentation step, using these files. We have performed additional extensive validation (as well as the one reported in this article) to ensure that the toolbox works correctly for both absolute perfusion images (CBF) and for perfusion-weighted difference images in which CBF computation is required.

### 3.2. Evaluation of hypoperfusion differences in healthy subjects at risk of developing Alzheimer's disease using the toolbox

Figure 4 shows the result of the rigid co-registration step between the 3D T1 weighted images and the CBF maps. Both images match the same anatomical space; the 3D T1 structural image has been re-sampled (as well as the tissue probability maps) to the low-resolution of the CBF map with a 'b-spline' interpolation method. Figure 5 shows the tissue probability maps of GM and WM for one subject.

The partial volume effect in ASL data is shown in Figure 6.  First and second rows show the 3D T1 weighted axial, sagittal and coronal planes for one patient and a detail of the left hippocampus and parahippocampal gyrus in the same planes respectively. Partial volume effect in the CBF map is shown in the third row and fourth row shows the result of Asllani's correction with a 5x5x1 low-resolution kernel. Figure 6 shows how the blood flow is increased_ in the whole region after PVC with Asllani's method. Table 1 shows quantitatively how the perfusion values change after the PVC in the whole region. Also, Table 2 shows the comparative results of CBF perfusion values for one subject after the PVC in the different subcortical brain structures for both hemispheres, showing the increase of perfusion for all ROIs, obtained after the PVC correction with Asllani's method.

The tissue probability maps (shown in Figure 5), the PVC method and the normalisation maps obtained from the anatomical images can be applied to the perfusion maps in order to obtain PVE corrected CBF maps in MNI space (Figure 7).

After performing each of the steps previously shown, in our cohort of 50 elderly subjects, the statistical group analysis was performed by means of a two-sample t-test in both cases: CBF maps with Asllani's PVE correction and the original CBF maps without PVE correction. Age and gender were introduced as confounding variables in the model. Figure 8 shows the results of these two analysis (Figure 8a.- PVE corrected, and 8b.- PVE uncorrected) for a family-wise error (FWE) corrected $p_{FWE}$<0.05 (cluster region of 300 voxels). Table 3 shows the T score and p values for the two analysis.

These results indicate decreased perfusion in healthy subjects at risk of developing Alzheimer's disease. Areas of significant hypoperfusion in Figure 8a (PVE-corrected) correspond to: caudate, hippocampi, thalamus, parahippocampal gyrus, amygdala, cingulate gyrus, precuneus, left and right insula, superior temporal lobe, uncus and choroid plexus.  Results from the statistical analysis without PVE correction (Figure 8b) show regions which appeared previously in the PVE corrected version (caudate, left hippocampus,  right thalamus, anterior cingulate, right insula and choroid plexus). In both analyses, part of the perfusion deficit appears displaced into the region of the ventricular space, probably because of the inherent blurring of the 3D FSE stack-of-spiral readout of the ASL pulse sequence employed in this study.  In a separate analyses, we confirmed that in fact the 'at-risk' cohort exhibits ventricular enlargement and reduction of grey matter volume in the vecinity of these areas.  The combination of these results with the inclusion of PV correction in ASL studies, forms part of a larger separate investigation which is beyond the scope of this paper.  These results are consistent with regions found by our studies and those of other authors [18,24,30-32].

## 4. Discussion

In this work, we have developed a MATLAB toolbox (ASAP) for systematically and automatically processing ASL datasets with minimal user intervention. The key advantage of ASAP is that it automates all the processing steps of ASL datasets for any number of subjects and the ability to work with reduced user input minimises the possibility of random and systematic errors. ASAP offers easily selectable option for almost all the stages of that process.The toolbox can produce perfusion data that is ready for statistical group analysis. A fully automated pipeline makes the data processing efficient and reduces potential mistakes by avoiding manual processing of individual steps. Besides, ASAP has a very friendly and easy to use GUI (Figure 2), allowing users to select the preferred options for each case. Depending on the datasets, users may change the options of some processing steps to optimize the processing quality.  Prior programming knowledge is not required. One limitation of other existing toolboxes lies in the requirement of programming knowledge, which limits their accessibility to users with programming skills.

In the present study, we applied ASAP to study possible changes in perfusion in a sample of healthy subjects in risk of developing AD. The analyses were run on a Macintosh OS X (10.6 Snow Leopard) computer with 8 GB of RAM and a 3.06 GHz Intel Core 2 Duo processor. The total running time was 3.44 hours (4.13 minutes per subject). The automatization of the whole post-processing pipeline minimises the variability introduced by human errors and decreases enormously the time needed to manually process all subjects. ASAP provides the images ready to perform statistical assessment. We have presented an evaluation of hypoperfusion in healthy subjects at risk of developing Alzheimer's disease and the results are consistent with those of previous studies that find decreased perfusion in Alzheimer's patients in similar regions. As an example, figure 8a and 8b show how the absence of PVE correction can lead to false negative findings. Regions affected by hypoperfusion have higher statistical significative and thus regions are more extensive due to the PVE correction. These results also highlight that PVE correction is required to maximise the predictive value of ASL in this field of research. Hypoperfusion in the right inferior insula and superior temporal lobe region can be detected with PVE corrected images whereas PVE uncorrected images cannot.This region in particular, is affected by atrophy in those subjects at the very early stages of Alzheimer's disease, but thanks to this PVE technique, it can be shown that a hypoperfusion pattern is prior to GM atrophy.

We envisage the intuitive and user friendly nature of the ASAP toolbox will help to facilitate the application of ASL in the clinical environment, where the method can be easily employed by clinicians and technicians without the need of intensive training or knowledge of image processing techniques. As mentioned earlier, processing data with ASAP can be very flexible and users have the freedom to design the most appropriate pipeline targeted to their data. In order to know which is the best pipeline for the user's data, there are important recommendations in the user's manual for a proper use of the different options available in the toolbox. There is a description of which is the best choice for each stage, depending on the input data, such as T1-weighted or T2-weighted structural scan, or in which cases is useful to apply a specific option, like the rough skull-stripping or the PVE correction.

High quality of co-registration between anatomical and perfusion images are key for optimal partial volume effect correction and normalisation steps. The results of employing our toolbox in this sample study, demonstrate that the software can produce a high level of accuracy of spatial normalisation, paying no penalty in quality as a result of fully automated operation. The assessment of the normalisation quality has been made qualitatively by comparing a minimum of 4 external cortical landmarks on the CBF maps and ensuring that they correspond to the same landmarks on the chosen template within a $\pm 3$mm range. The same assessment has been made with at least 2 other sub-cortical landmarks. Nevertheless, further improvements such as the use of higher field warping options as those of the DARTEL library of SPM [33] are currently being incorporated for a subsequent version. Although this option will require the selection of a subgroup of images to generate an intermediate group specific template and computational time is likely to increase, the method may be more accurate and precise for group comparisons.

One of the disadvantages of automatization is that some mistakes might go undetected. To ameliorate this problem, ASAP's interface includes a "Quick check" option for displaying the resultant MNI-normalized ASL images in a web browser once the processing is completed. This option allows an convenient quality assurance method, making it possible to check the normalization quality or whether any intermediate step has failed. Many functionalities and features are open for improvement in future versions of the software. Other labelling schemes as well as multi-TI ASL sequences, will be included in further versions of ASAP.

## 5. Conclusion

In conclusion, the results for this specific study show the applicability of ASAP in the study of perfusion changes in elder people at risk of developing AD. Furthermore, these clinical results are consistent with previous AD studies. In summary, our toolbox provides a simple, flexible and reliable solution for ASL-related studies. It has an extendable design, and new functions or utilities can and will be added in the future. The ASAP manual and software can be obtained freely at *sites.google.com/site/asltoolbox*. Feedback from users will be encouraged to ensure the updated of the ASAP toolbox, in order to include future improvements in image processing methodology. We hope for rich participation from the ASL community.

**Acknowledgments**: ASAP was partially supported by the COST Action "Arterial Spin Labelling Initiative in Dementia" (BMBS COST Action BM1103).

## Tables

|  | Original CBF | PVC CBF |
|---|---|---|
| Left Hippocampus | 40±10 | 46±9 |
| Right Hippocampus | 42±11 | 42±11 |
| Left Parahippocampal Gyrus | 40±10 | 48±10 |
| Right Parahippocampal Gyrus | 35±8 | 44±10 |

Table 1. CBF perfusion values (ml/100g/min) in the left and right hippocampus and parahippocampal gyrus (same regions and patient that Figure 6). Left column shows the original CBF values (mean±std) and right column shows the CBF values (mean±std) after the PVC using the Asllani's method with a 5x5x1 low-resolution kernel.

|  |  | Original CBF | PVC CBF |
|---|---|---|---|
| Left Hemisphere | Amygdala | 35±10 | 37±7 |
|  | Caudate | 35±13 | 38±10 |
|  | Hippocampus | 34±10 | 38±9 |
|  | Pallidum | 38±11 | 49±12 |
|  | Putamen | 42±8 | 44±6 |
|  | Thalamus | 44±17 | 54±16 |
| Right Hemisphere | Amygdala | 34±7 | 38±4 |
|  | Caudate | 31±15 | 35±11 |
|  | Hippocampus | 34±12 | 38±8 |
|  | Pallidum | 27±6 | 42±11 |
|  | Putamen | 41±8 | 45±5 |
|  | Thalamus | 44±19 | 54±18 |

Table 2. Example of CBF values (ml/100g/min) in the different subcortical brain structures for both hemispheres. Left column shows the original CBF values (mean±std) while right column shows the CBF values (mean±std) after the PVC using the Asllani's method with a 5x5x1 low-resolution kernel.

|  | T value (p-value) Non-PVC | T value (p-value) PVC |
|---|---|---|
| Left Hippocampus | 2.58 (0.006465) | 3.28 (0.000958) |
| Right Hippocampus | 2.70 (0.004746) | 3.25 (0.001045) |
| Left Parahippocampal Gyrus | 1.76 (0.042325) | 2.24 (0.01483) |
| Right Parahippocampal Gyrus | 1.66 (0.051651) | 2.52 (0.007522) |

Table 3. T scores and p-values (in brackets) of the statistical group analysis ($p_{FWE}<0.05$, cluster region of 300 voxels) in the same regions that Table 1 and Figure 6. Left column shows the results for the original CBF maps (non-PVC corrected) and right column shows the results for the final CBFafter the PVC using the Asllani's method with a 5x5x1 low-resolution kernel.

## Figure Legends

Figure 1. Pipeline for processing ASL datasets in ASLToolbox. Each box represent a main step in ASLToolbox's procedure and top dotted line boxes represent the input data.

Figure 2. Graphical User Interface of the ASLToolbox for loading dataset. It consists of three main sections namely "Input Files", "Output Directories" and "Options".

Figure 3. Graphical User Interface for ROI Statistics analysis in ASLToolbox.

Figure 4. Example of the result of the co-registration step between a 3DT1 weighted image (background) and a CBF map (overlay). The 3DT1 structural image has been re-sampled to the resolution of the ASL data.

Figure 5. Tissue probability maps of GM (rows 1,3,5) and WM (rows 2,4,6) of a subject, registered onto the 3DT1 structural scan on a sagital (rows 1,2), coronal (rows 3,4) and axial planes (rows 5,6). These maps were used for the PVC of the CBF maps.

Figure 6. 3D T1 weighted axial, sagittal and coronal planes for one patient (first row), detail of the left hippocampus and parahippocampal gyrus in the same planes (second, third and fourth rows). Color overlay in third and fourth row correspond to the CBF map. Third row shows the partial volume effect in the original CBF and fourth row shows the result of Asllani's PVC with a 5x5x1 low-resolution kernel.

Figure 7. Example of a final smoothed (6mm Gaussian kernel), MNI normalised and PVE corrected CBF map as an overlay onto the 3DT1 MNI template.

Figure 8. Results of the statistical group comparison: significative hypoperfusion regions for healthy subjects in risk of developing AD pFWE<0.05 (minimum cluster 300 voxels) for: (a) PVE corrected CBF maps and (b) CBF maps with no PVE correction